\documentclass{emulateapj}
\usepackage{subfigure}
\shorttitle{SN and LGRB Locations in Their Host Galaxies}

\shortauthors{Kelly et al.}
\begin{document}

\title{Long $\gamma$-ray bursts and Type Ic core-collapse supernovae have similar locations in hosts}

\author{Patrick L. Kelly\altaffilmark{1,2}}

\affil{Kavli Institute for Particle Astrophysics and Cosmology, Stanford University, 382 Via Pueblo Mall, Stanford, CA 94305-4060}
\email{pkelly3@stanford.edu}
\and
\author{Robert P. Kirshner\altaffilmark{2}}
\affil{Harvard-Smithsonian Center for Astrophysics, 60 Garden St., Cambridge, MA 02138}
\email{rkirshner@cfa.harvard.edu}
\and
\author{Michael Pahre\altaffilmark{2}}
\affil{Harvard-Smithsonian Center for Astrophysics, 60 Garden St., Cambridge, MA 02138}
\email{mpahre@cfa.harvard.edu}

\altaffiltext{1}{Kavli Institute for Particle Astrophysics and Cosmology, Stanford University, 382 Via Pueblo Mall, Stanford, CA 94305-4060}
\altaffiltext{2}{Harvard-Smithsonian Center for Astrophysics, 60 Garden St., Cambridge, MA 02138}

\keywords{supernovae: general --- gamma rays: bursts}

\begin{abstract}
When the afterglow fades at the site of a long-duration $\gamma$-ray burst (LGRB), Type Ic supernovae (SN Ic) are the only type of core collapse supernova observed. Recent work found that a sample of LGRB in high-redshift galaxies had different environments from a collection of core-collapse environments, which were identified from their colors and light curves.  LGRB were in the brightest regions of their hosts, but the core-collapse sample followed the overall distribution of the galaxy light. Here we examine 504 supernovae with types assigned based on their spectra that are located in nearby (z $<$ 0.06) galaxies for which we have constructed surface photometry from the Sloan Digital Sky Survey (SDSS).  The distributions of the thermonuclear supernovae (SN Ia) and some varieties of core-collapse supernovae (SN II and SN Ib) follow the galaxy light, but the SN Ic (like LGRB) are much more likely to erupt in the brightest regions of their hosts. The high-redshift hosts of LGRB are overwhelmingly irregulars, without bulges, while many low redshift SN Ic hosts are spirals with small bulges. When we remove the bulge light from our low-redshift sample, the SN Ic and LGRB distributions agree extremely well. If both LGRB and SN Ic stem from very massive stars, then it seems plausible that the conditions necessary for forming SN Ic are also required for LGRB. Additional factors, including metallicity, may determine whether the stellar evolution of a massive star leads to a LGRB with an underlying broad-lined SN Ic, or simply a SN Ic without a $\gamma$-ray burst.
\end{abstract}

\maketitle
\begin{figure*}[ht!]
\centering
\plottwo{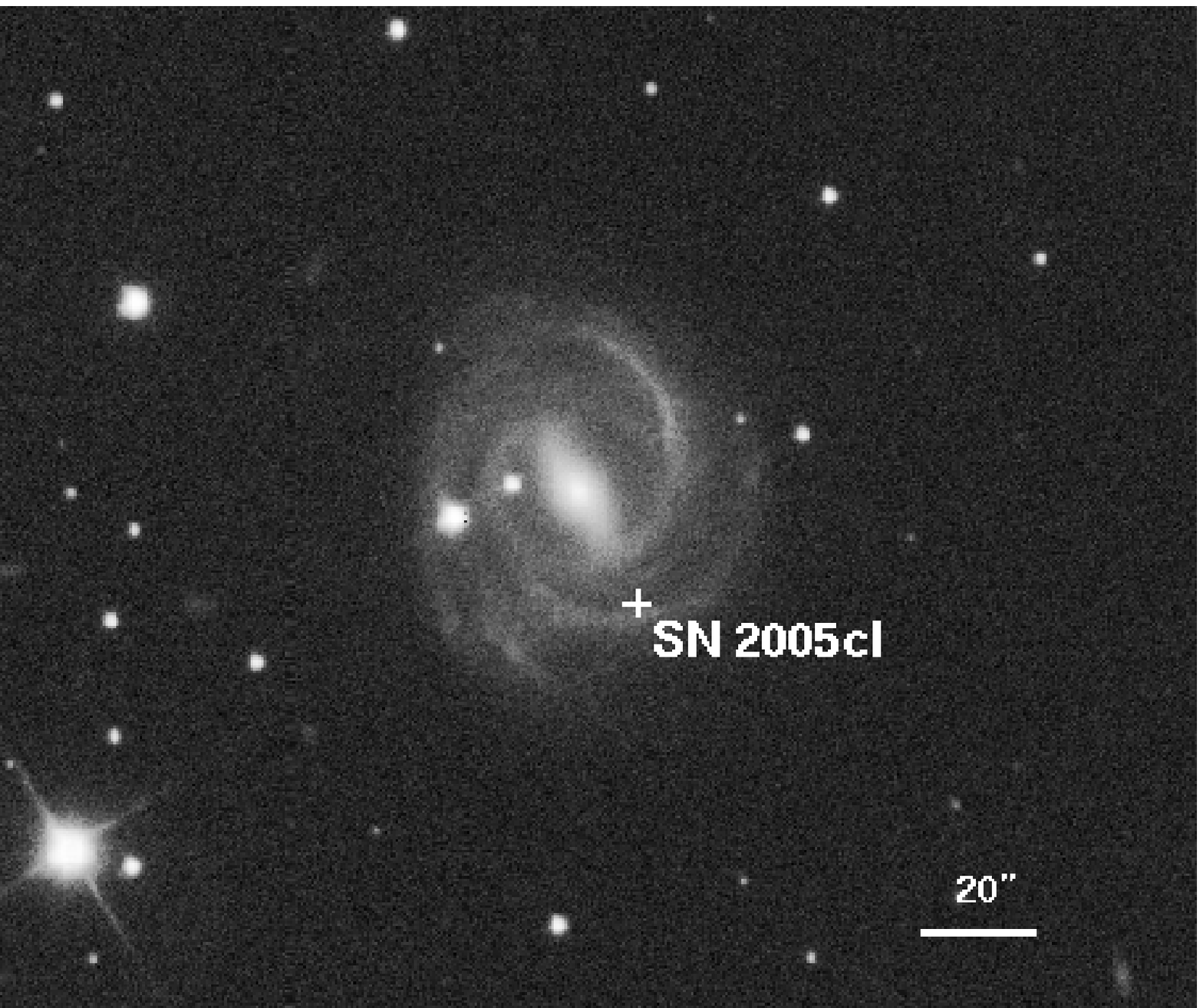}{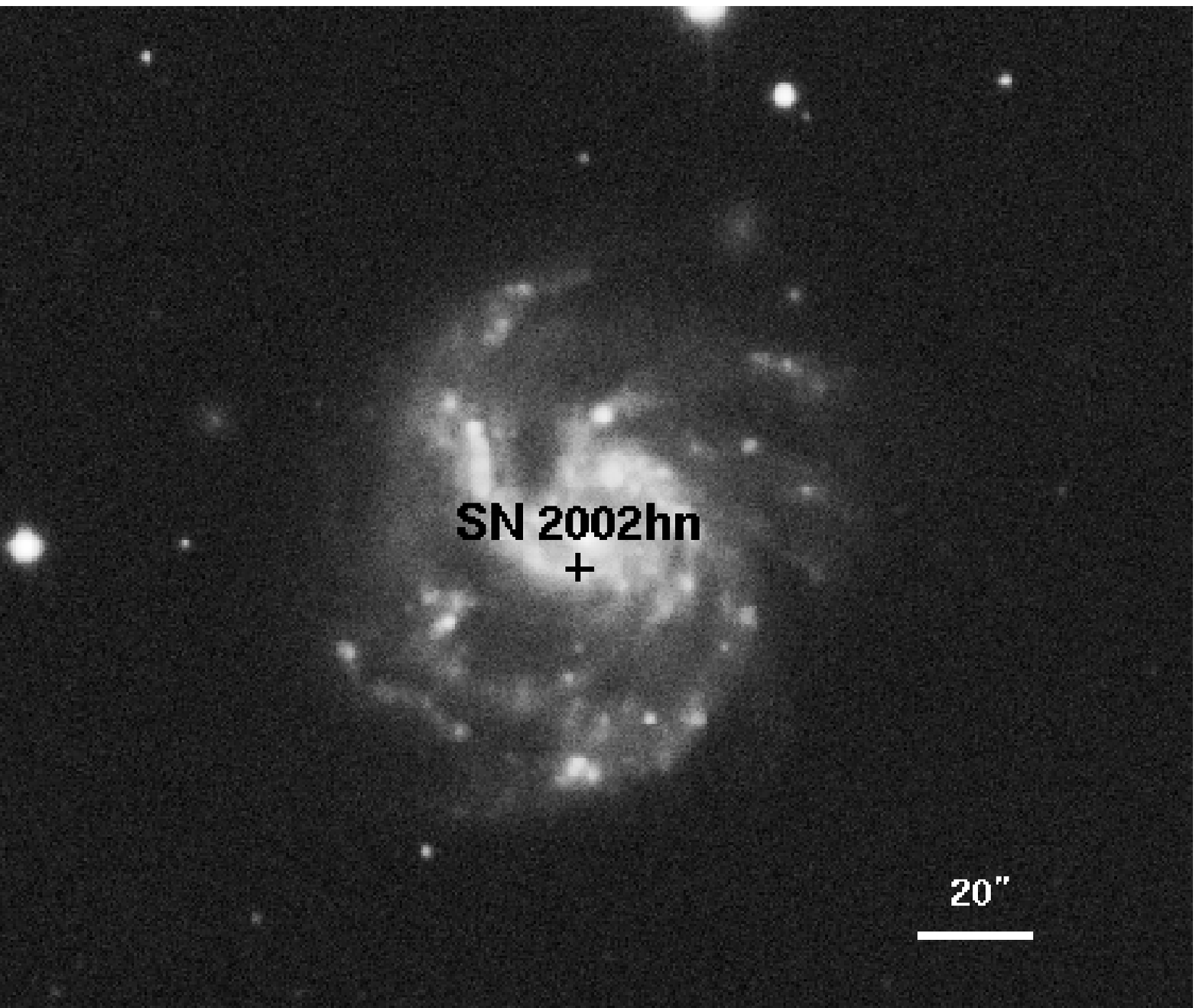}
\caption{Typical galaxies in the sample. (a) SDSS \textit{g'}-band image of size $3.4 \times 3.7 $~arcmin$^2$ of the host galaxy M-01-53-20 for SN~2005cl, a SN IIn. (b) SDSS \textit{g'}-band image of size $3.4 \times 3.7 $~arcmin$^2$ of the host galaxy NGC~2532 for SN~2002hn, a SN Ic. }
\label{fig:example}
\end{figure*}
\section{Introduction}
Type Ic supernovae (SN Ic) are defined by the absence of hydrogen, helium, and strong silicon features in their spectra.  They are the only type of supernova observed after the fading of the afterglow from long-duration $\gamma$-ray bursts (\citealt{ga98}; \citealt{ma03}; \citealt{st03}; \citealt{hj03}; \citealt{mod06}).  SN Ic represent the extreme in a progression of increasing mass loss along the path to core collapse, ranging from SN II, which have hydrogen in their spectra \citep{fi97}, through SN Ib, whose spectra lack hydrogen but show helium lines, to SN Ic which have neither hydrogen nor helium in their spectra. Two long-duration $\gamma$-ray bursts (LGRB) and an event spectrally similar to but less energetic than an LGRB, an X-ray flash (XRF), have been firmly connected to coincident SN Ic, all of which had ``broad-lined" spectral features suggestive of ejecta velocities on the order of 30,000 km s$^{-1}$. SN 1998bw was observed at the same position as a coincident LGRB \citep{ga98}; in the case of SN 2003dh, a residual SN Ic spectrum was discovered after subtracting a power-law continuum from the LGRB spectrum (\citealt{ma03}; \citealt{st03}; \citealt{hj03}), and a SN Ic spectrum was clear in the follow-up spectra to the XRF counterpart to SN 2006aj (\citealt{mod06}; \citealt{mi06}; \citealt{sol06}; \citealt{pi06}). Although our investigation is not tied to any particular theoretical picture, the qualitative features of ``broad-lined" SN Ic are consistent with the collapsar picture, which attributes LGRB to tightly collimated jets that emerge during core-collapse SN explosions \citep{wo93}. In this case, the SN Ic spectrum suggests that the star has shed, or burned through deep mixing, much of its hydrogen and helium, making jet escape less difficult, while the broad lines in the spectrum suggest that there are high velocities impressed on the remaining gas. 

Broad-lined SN Ic constitute a small fraction of the observed SN Ic population (\citealt{pod04}; \citealt{gu07}). While some broad-lined SN Ic are exceptionally luminous, there are examples, such as SN 2002ap ($E \sim$ (4-10) $\times 10^{51}$ ergs), which match best with a low-energy model \citep{ma02}. SN 2006aj, the broad-lined SN Ic connected to an XRF, had an energy only comparable to core-collapse SN \citep{ma07}.  Because LGRB jets are inferred to be highly collimated, some fraction of broad-lined SN Ic could be LGRB whose jets point in a direction that does not include Earth.  From observations of radio emission that is assumed to be isotropic, the hypothesis that all broad-lined SN Ic have off-axis jets was ruled out at the 84\% confidence level \citep{so06}, suggesting that a significant fraction of the broad-lined SN Ic have no accompanying relativistic jet. Metallicity may be an important factor in determining whether a massive star produces a LGRB: in a small, but growing, sample, broad-lined SN Ic without an associated LGRB were found to inhabit more metal-rich galaxies than broad-lined SN Ic associated with LGRB  \citep{mod08}.

\section{Data}

The study of LGRBs has used the locations of events in high redshift hosts to good effect (\citealt{bl02}; \citealt{fr06}). In this paper, we use a technique developed by \citealt{fr06} (F06) for use on distant LGRB host galaxies, applying it to the low-redshift SN host sample to establish the spatial distribution of supernova progenitors in this less exotic setting. Low-redshift SN Ib/SN Ic with host galaxies in the Sloan Digital Sky Survey (SDSS) were recently found to have more metal-rich hosts and occur at somewhat smaller offsets from their hosts' centers than SN II \citep{pr08}. F06 compared the positions of LGRBs to ``core-collapse" SN discovered by the GOODS ACS Treasury Program (\citealt{ri04}; \citealt{st04}), with the mean redshift of LGRBs of 1.25 and a mean redshift of 0.63 for the core-collapse supernovae. The Higher-Z team generally did not observe further any SN that did have not the colors of SN Ia. These rejects for cosmology (``Not SN Ia") constitute the F06 core-collapse sample. F06 found that the core-collapse SN defined this way follow the light distributions of their hosts: the probability of a SN occurring in a given pixel is proportional to the counts in that pixel. LGRB positions, in contrast, were concentrated at the regions of highest surface brightness in their hosts.  This is why F06 concluded that LGRB and core-collapse SN have different environments. In our nearby sample, we have the advantage of more complete spectroscopic information, which permits a more refined sorting by supernova type.  Unlike F06, we find a strong resemblance between SN Ic and the LGRB positions in their hosts.

Our sample was drawn from 4603 SN discoveries reported to the International Astronomical Union (IAU) through 2008 March 3, collected in the Asiago Supernova Catalog \citep{ba99}. We selected SN in hosts occurring inside SDSS DR6 coverage with $z < 0.06$, to allow for high signal-to-noise photometry and host images with sizes significantly larger than the seeing. Typical galaxies in the sample, M-01-53-20 and NGC~2532, and the locations of SN 2005cl (Type IIn) and SN 2002hn (Type Ic) are shown in Figure~\ref{fig:example}.

\subsection{SN Classifications}
The identification of SN spectroscopic types (see \citealt{fi97} for a review) has improved in the last 15 years with the use of cross-correlation between each candidate spectrum and existing templates. Yet even with cross-correlation, identification can be imperfect for late-time or low signal-to-noise ratio spectra. Mistakes can occur such as in the case of SN 2004aw where reports suggested that it was a SN Ia \citep{be04} but was eventually detemined to be a SN Ic by \citealt{fi04}. In this Paper, we require the SN used in the analysis to have types unambiguously determined from spectra, so SN with split classifications such as SN Ib/Ic, types determined from light curves, or with uncertain classifications were not used.  In our data table Table~\ref{tab:data}, we also present measurements for SN whose classifications do not meet this criterion. Each SN type plotted in this Paper includes all subtypes and SN with peculiar (``pec") designations. These subtypes include broad-lined (``Ib-bl") for SN Ic and linear (``IIL"), plateau (``IIP"), with narrow emission lines (``IIn"), and the transitional type (``IIb") for SN II.

\begin{deluxetable}{lcccccc}[!b]
\tablecaption{Sample Construction}
\tablecolumns{6}
\tablehead{\colhead{Criterion}&\colhead{Ia}&\colhead{Ib}&\colhead{Ic}&\colhead{II}&\colhead{All}}
\startdata
Asiago/Velocities/Type&2057&77&145&1051&3285&\\
Vel.$<$0.06/Position&870&71&125&890&1883&\\
SDSS DR 6&360&24&47&361&816&\\
Residual SN Light&256&18&32&238&561&\\
Stellar Contamination&243&18&30&223&540&\\
Host Too Large&239&18&30&223&527&\\
Spectroscopic Confirmation&234&18&27&214&508&\\
Confident Classification&233&18&26&213&505&\\
Host Detected &232&18&26&213&504&\\
\enddata
\tablecomments{Number of SN of each spectroscopic type remaining after applying each inclusion criterion. (1) SN collected in the Asiago Catalog updated through March 3, 2008 with redshift and type information; (2) hosts with z $<$ 0.06 and SN position coordinates in the host galaxy; (3) inside SDSS DR 6 coverage; (4) hosts observed when expected residual SN light is insignificant; (5) no contamination from bright stars; (6) host not too large to effectively process; (7) classification from spectrum not light curve; (8) lack of ambiguity in spectroscopic classification; (9) host detected. SN 2007qc was the only SN without a detected host. }
\label{tab:selection}
\end{deluxetable} 
\subsection{Imaging}
Table~\ref{tab:selection} lists the criteria we applied to construct the sample and the number of each remaining after each cut. Although we assembled mosaics for galaxies that lay on the CCD chip edges, 13 SN  host galaxies (SN 1909A, SN 1951H, SN 1970G, SN 1971I, SN 1981K, SN 1989M, SN 1993J, SN 1997bs, SN 2001R, SN 2003cg, SN 2004am, SN 2006X, SN 2008ax) were so large that they were not used because the SDSS pipeline has difficulty with photometry in frames dominated by large galaxies. SN in the Asiago Supernova Catalog are reported with an associated host galaxy, and that host galaxy identification was assumed to be correct. The images of 21 SN host galaxies included nearby bright stars that were impossible to mask out (SN 1980D, SN 1986E, SN 1990ag, SN 1998B, SN 1999be, SN 2003U, SN 2003W, SN 2003eh, SN 2004co, SN 2004dt, SN 2005bk, SN 2005la, SN 2005ms, SN 2006ac, SN 2006at, SN 2007O, SN 2007aa, SN 2007cl, SN 2007sa, SN 2007sp, SN 2008E). We could not include SN 2007qc because its host was not detected.

The SDSS \citep{yo00} DR 6 \citep{ad08} includes 9583 square degrees of 53.9-second imaging in the Sloan filter set (\textit{u'},\textit{g'},\textit{r'},\textit{i'},\textit{z'}) on a wide-field 2.5 m telescope in Apache Point, New Mexico. The Sloan filters partition the spectrum from the near-infrared detector sensitivity limit to the ultraviolet atmospheric cutoff in non-overlapping bands \citep{fu96}. Individual Sloan frames have a 2048 x 1498 array of 0.396$^{\prime\prime}$ square pixels, creating a 13.5$^{\prime}$ $\times$ 9.9$^{\prime}$ field. The mean seeing in our \textit{g'}-band observations was 1.41$^{\prime\prime}\pm$0.24$^{\prime\prime}$.

\section{Methods}
\subsection{Fractional Flux} 

Our goal is to make measurements of the low-z sample to compare with the data presented by F06. We measure the surface brightness (1) in the pixel at the SN position and (2) by averaging inside a 0.4 kpc aperture centered at the SN position. We used SDSS \textit{g'}-band images for comparison because the \textit{g'}-band registers the spectral regions where galaxies emit most of the light detected by F06. Source Extractor \citep{bert96} was used to find the set of pixels with the host galaxy light by selecting contiguous pixels with signal-to-noise ratios greater than 1, identified after applying a small pyramidal convolution mask to the image. From the distribution of pixel values in the galaxy light distribution and the local SN surface brightness measured in the pixel at the SN location, we calculate the ``fractional flux." The fractional flux is the sum of counts registered in all pixels with fewer counts than measured at the SN location divided by the sum of all counts associated with the galaxy, 

\begin{eqnarray}
\lefteqn{\textrm{fractional flux} = } \nonumber \\
& & \frac{\sum_{\textrm{\scriptsize{counts in pixel}} < \textrm{\scriptsize{counts at SN}}} \textrm{counts in pixel} } {\sum_{\textrm{\scriptsize{all galaxy pixels}}}\textrm{counts in pixel}}.\label{eqtn:fracflux}
\end{eqnarray}

This simple statistic allows a direct measurement of how SN events are distributed over their hosts. For galaxies with bulges, we make an additional calculation of the fractional flux after removing bulge light by replacing pixels values inside a circular region encompassing the bulge-dominated center with the mean of the perimeter pixel values. 


\subsection{Residual SN Light}

From representative light curves \citep{cap01}, we found that absolute magnitudes decay to fainter than -5 Johnson V magnitudes within one year for SN Ia, II except IIn, and Ib/Ic. We excluded SN when the SDSS observation of the host was made in the time period ranging from 3 months before the supernova report through 24 months after detection. SN IIn sometimes show very slow luminosity evolution, so to avoid contamination, we excluded SN hosts with Sloan data taken over the range 2 years before detection to 5 years after the SN discovery. Converting SDSS magnitudes to Johnson \textit{V} band using kcorrect (v. 4.13), a SED fitting program developed for use with the SDSS filter set \citep{bl03}, we found an absolute Johnson \textit{V} magnitude of galaxy light in the pixel at the SN position of -8.81$\pm$2.04 for SN satisfying the previous time-selection criteria, roughly a factor of 30 times brighter than bright supernovae.  These rules put us in a good position to avoid contamination of the galaxy surface brightness by light from the supernova.                                        

\section{Results}
\begin{figure*}[htp!]
\centering
\plotone{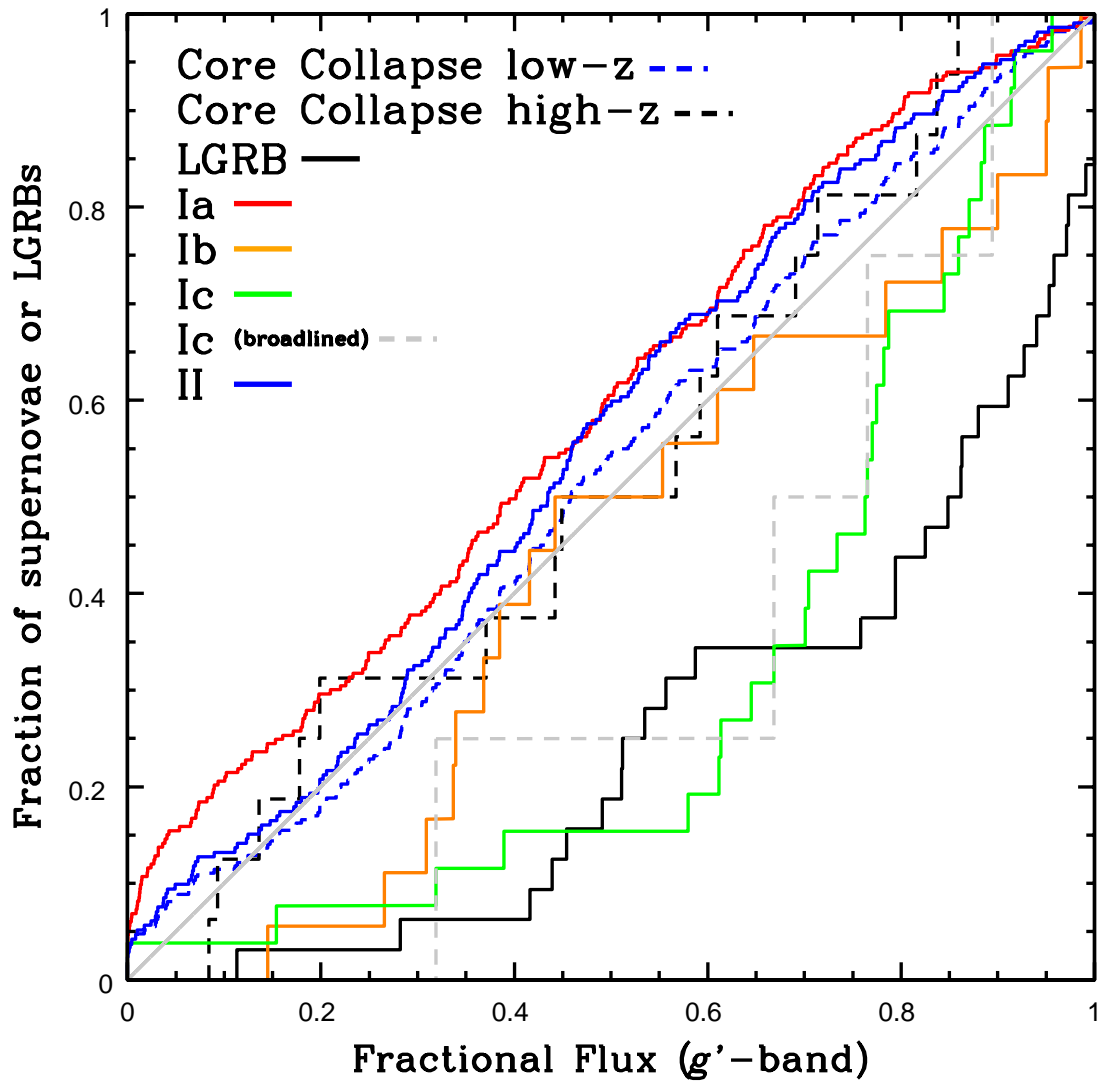}
\caption{Low-redshift SN \textit{g'}-band and high-redshift core-collapse SN and LGRB (from \citealt{fr06}) fractional flux distributions. SN Ic (N=26) (including broad-lined) are in brighter regions of their hosts than SN Ib (N=18; p=0.033) and SN II (N=213; p=2x10$^{-6}$). SN Ic are absent from the top 3\% of galactic flux fraction distributions, which is likely due to the presence of bright central bulges and further explored in Figure~3. The SN Ib (N=18) distribution has an 5\%, SN Ic distribution a 6\%, and SN II (N=213) distribution only a 6x10$^{-5}$ \% probability of being drawn from the same set as LGRBs. The high redshift core-collapse sample, likely consisting predominantly of SN II, and the low redshift core-collapse population from our sample are highly similar (p = 0.93). SN Ia (N=232) show some similarity to the SN II and the core-collapse (N=16) distributions, with p=0.22 and p=0.63 respectively. SN Ib and a population tracing the light of the hosts have p=0.36.}
\label{fig:pixstat}
\end{figure*}

We plot the cumulative distribution of the  \textit{g'}-band ``fractional flux" statistic in Figure~\ref{fig:pixstat}: The Y-axis indicates the fraction of the supernova population with ``fractional flux" values less than the X-axis value. If the population follows the light distribution, the fractional flux plot will be just a straight line connecting (0,0) and (1,1). 

\begin{figure*}[htp!]
\centering
\plotone{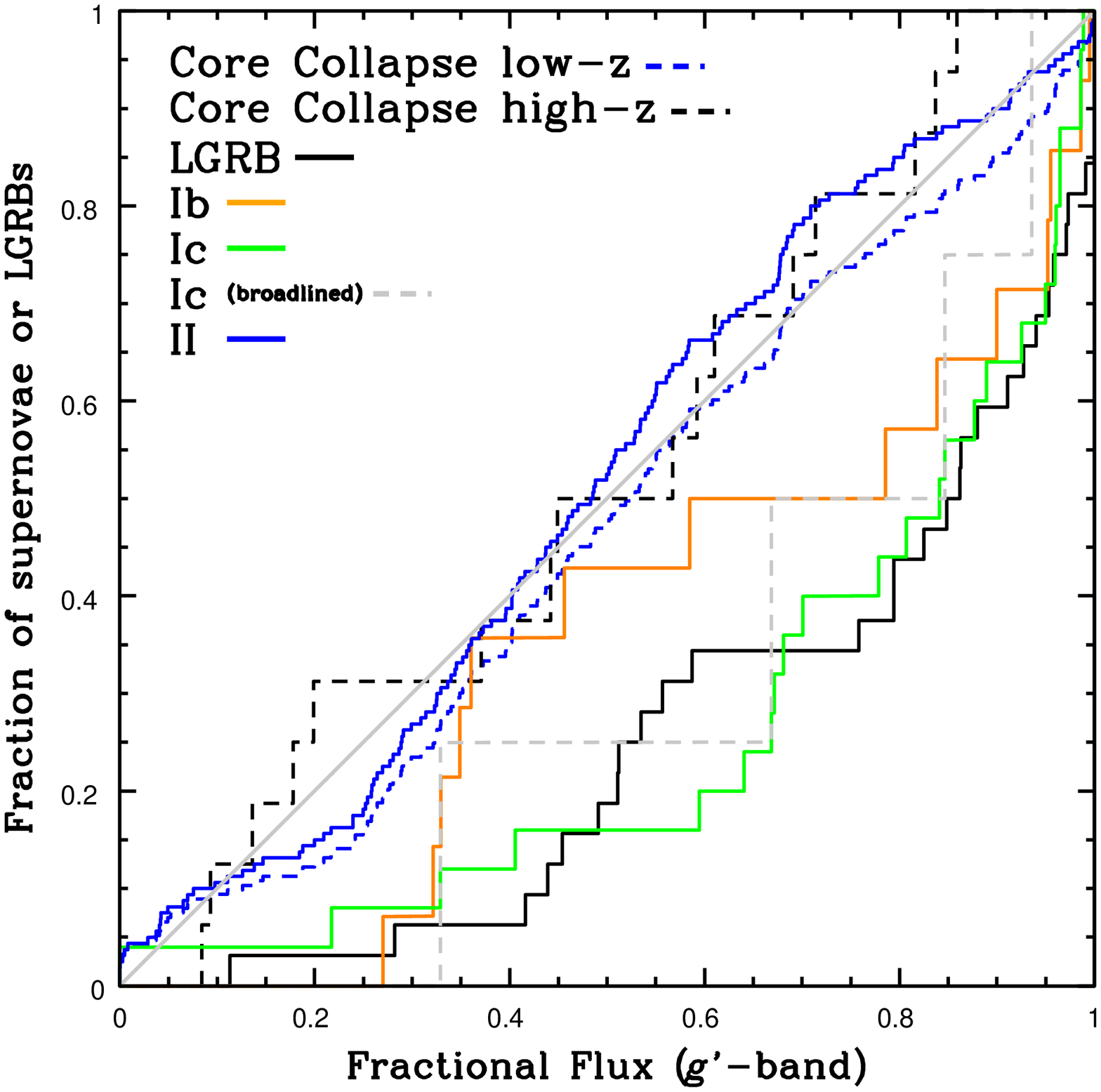}
\caption{Bulge-subtracted low-redshift SN \textit{g'}-band and high-redshift core-collapse SN and LGRB (from \citealt{fr06}) fractional flux distributions. After removing the bulge light that is present in low-redshift SN hosts but not the high-redshift and irregular LGRB hosts, there is a high probability (p=0.66) that the SN Ic (N=25)(including broad-lined) and the LGRB (N=32) distributions are drawn from the same set. In contrast, it is highly unlikely  that SN II (N=160) and LGRB are drawn from the same set (p=4x10$^{-6}$). Bulge subtraction in SN II hosts does not strongly affect the SN II distribution, which remains linear and in good agreement with high-z core-collapse SN (N=16; p=0.76). There is a 30\% probability that the SN Ib (N=14) distribution is identical to the LGRB distribution and a 20\% probability that it is identical to the Ic distribution. }
\label{fig:pixstat2}
\end{figure*}

\subsection{Locations in Host Light}
As can be seen in Figure~\ref{fig:pixstat}, SN Ic are much more likely than SN Ib and SN II to be found in the brightest regions of their host galaxies.  The distribution of the four ``broad-lined" SN Ic in this Figure is consistent with the distribution of the larger SN Ic population. A Kolmogorov-Smirnov (KS) test finds that broad-lined SN Ic are more likely to be drawn from the same population as SN Ic (N=26; p=0.96) than SN II (N=213; p=0.16). Broad-lined SN Ic are the type found at the sites of LGRB, but we have too few cases in our sample to say whether they track that distribution more closely than other SN Ic.  In contrast to the strongly skewed SN Ic, the SN Ia, SN Ib, and SN II all have approximately linear distributions, indicating that the distribution of supernovae is the same as the distributions of light in their hosts. A KS test finds a very low probability (p = 2x10$^{-6}$) that SN Ic  and SN II  are drawn from the same population, and low probability (p = 0.033) that SN Ic are drawn from the same population as SN Ib (N=18). The distinction between the SN Ic and SN II distributions provides an interesting contrast to work by \citet{vd96}, which found that samples of 18 SN Ib/c and 32 SN II had similar associations with nearby {\ion{H}{2}} star-forming regions.  

Figure~\ref{fig:pixstat} compares our \textit{g'}-band fractional flux results for low redshift SN, all of which have spectroscopic types, to the results reported by F06 at high redshift. Both LGRB and SN Ic samples are concentrated in the bright regions of their hosts, in sharp contrast to SN II.  We see good agreement between the ``core-collapse" SN (N=16) from F06, which probably consists primarily of SN II, and our low-redshift core-collapse SN (N=271; p=0.93) as well as low-redshift SN II (N=213; p = 0.74). If the observed ratio of (SN Ib + SN Ic)/SN II seen at low redshift is similar to that in the GOODS sample, then $\sim$1/5 of the core-collapse sample will be SN Ib or SN Ic, the fraction found in nearby surveys (\citealt{va91}; \citealt{ca99}). Furthermore, it is possible that a number of bright SN Ib and SN Ic were mistaken for SN Ia on the basis of their colors and luminosities. If the true connection of LGRB with core-collapse events is only with SN Ic, then this signal may well have been lost in the noise of a sample dominated by SN II.  Unlike the LGRBs in F06, SN Ic in our sample are absent from the very brightest pixels in their hosts, with no fractional flux values greater than 0.97.  This difference may arise because the brightest few pixels in the low-redshift supernova galaxies are dominated by bulge light. 

Despite the spectroscopic similarities between SN Ib and SN Ic, their progenitor populations track their hosts' light differently. In Figure~\ref{fig:pixstat}, the SN Ib distribution is more similar to the SN II distribution (p=0.36) than to the SN Ic distribution (p=0.033). A reasonable suggestion is that SN Ic occur in the brightest regions because they are the largest star-forming regions which produce the most massive stars. If so, SN Ib may have somewhat more moderate progenitor masses than SN Ic. While the very massive stars that experience Wolf-Rayet/Luminous Blue Variable phases lose their outer envelope through stellar winds, less massive stars may require a binary companion to strip the outer H layer. Whether a star not massive enough to explode as a SN Ic becomes a SN Ib or a SN II may depend in part on the presence or absence of such a binary companion.

\subsection{Locations in Host Disk Population Light}

Because high-redshift LGRB hosts have no bulge component while some low-redshift hosts do, we subtract the bulge from our low-redshift hosts and remeasure fractional flux values. Figure~\ref{fig:pixstat2} plots the fractional flux distributions for SN in hosts where we have removed bulge light by constructing a circular aperture encompassing the bulge and replacing enclosed pixel values by the mean of pixel values on the perimeter. We now find a 66\% probability that the LGRB (N=32) and SN Ic (N=25) distributions could be drawn from a single underlying population. It remains highly unlikely that SN Ic and SN II (N=160) are drawn from the same set (p=2x10$^{-5}$) or that SN II and LGRB are drawn from the same set (p=4x10$^{-6}$).

SN were only included in Figure~\ref{fig:pixstat2} if we could (1) visually identify and then subtract the bulge or (2) rule out a significant bulge. We therefore excluded bulge-dominated and edge-on galaxies but included SN with dwarf galaxy hosts. The absolute magnitudes of the dwarf galaxy hosts are listed in Table~\ref{tab:hosts}. A total of 14 of 18 SN Ib hosts and 25 of 26 SN Ic hosts either had their bulges removed or had no significant bulge and so were included in Figure~\ref{fig:pixstat2}. 

\begin{deluxetable}{lcc}[htp!]
\tablecaption{Dwarf galaxy absolute magnitudes}

\tablecolumns{3}
\tablehead{\colhead{SN}&\colhead{\textit{g'}-Absolute Magnitude}&\colhead{Type}}

\startdata
1995ah & -16.88 & II  \\ 
2004hx & -16.78 & II  \\ 
2007I  & -16.63 & Ic  \\ 
1998bm & -16.51 & II  \\ 
2005lb & -16.40 & II  \\ 
2007lj & -16.24 & II  \\ 
2007dq & -16.08 & II  \\ 
2007bu & -15.88 & II  \\ 
1997az & -15.57 & II  \\ 
2007em & -15.36 & II  \\ 
2007ld & -15.23 & II  \\ 
2006fg & -15.14 & II  \\ 
1998bv & -15.13 & II  \\ 
2007eh & -14.93 & II  \\ 
2005hm & -14.93 & Ib  \\ 
2005lc & -14.65 & II  \\ 
2007bg & -14.09 & Ic  \\ 
\enddata


\tablecomments{The faint absolute magnitudes of these host galaxies indicate that they are likely dwarf galaxies without a significant bulge, and we include them in Figure~\ref{fig:pixstat2}.}
\label{tab:hosts}
\end{deluxetable}

\begin{figure}[b!]   
\centering
\plotone{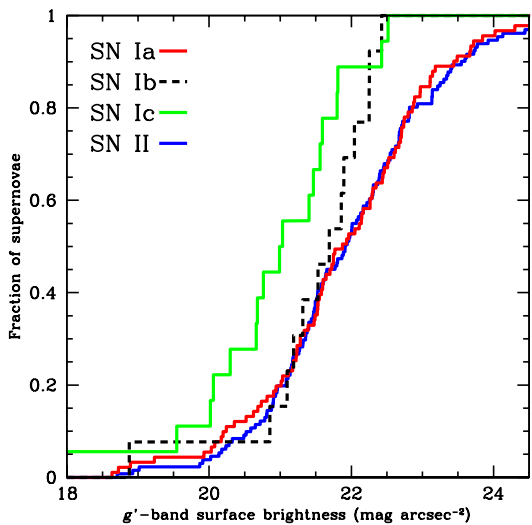}
\caption{Surface \textit{g'}-band brightnesses in 0.4 kpc apertures at the sites of SN.  This plot shows that the \textit{g'}-band surface brightnesses at SN Ic (N=18) sites are greater than those at SN Ib (N=13) and SN II (N=131) sites with p = 0.13 and p = 0.006. There is an 11\% probability that the SN Ib and SN II distributions are drawn from a single distribution. }
\label{fig:absspot}
\end{figure}

\subsection{Surface Brightness Measurements}

To complement fractional flux, which measures the brightness at the SN or LGRB location \emph{relative} to light across the entire host galaxy, Figure~\ref{fig:absspot} plots surface brightnesses in the \textit{g'}-band in a 0.4 kpc aperture at the SN location. To compute the K-correction, we required non-negative fluxes in each aperture in all five SDSS bands, which eliminated 44 of 234 SN Ia, 1 of 18 SN Ib, 2 of 26 SN Ic, 24 of 213 SN II.  We excluded measurements where the 0.4 kpc sampling aperture subtends an angle smaller than two pixels widths (0.8$^{\prime\prime}$), corresponding to a maximum recessional velocity of 7530 km s$^{-1}$. SN Ic (N=18) locations have higher surface brightnesses than SN Ib (N=13) and SN II (N=131), with relatively low probabilities (p=0.13 and p=0.005) that they are drawn from the same population.

All data plotted in this paper are collected in Table~\ref{tab:data}.

\section{Tests for Possible Systematic Effects}

To test for the possibility that the fraction of host light recovered using the signal-to-noise cutoff depends on the host luminosity, we created artificial disk galaxies with representative angular sizes and apparent magnitudes for our sample. With the exception of the undetected host galaxy of SN 2007qc, the faintest host was magnitude 21.0 (SN 2007aj) with 8 galaxies between magnitude 20 and 21. We do not find significant bias in the fraction of host light recovered for galaxies in our sample.  

To approximate the effects of increasing distance to host galaxies on surface brightness measurements in the 0.4 kpc aperture plotted in Figure~\ref{fig:absspot}, we convolved and resampled images of the 86 SN in our sample with recessional velocities less than 2000 km s$^{-1}$. Repeating surface brightness measurements to simulate recessional velocities to 10,000 km$^{-1}$, we find an insignificant trend toward fainter measured surface brightness of 0.02 mag per 1000 km$^{-1}$.

\section{Sample Comparison}
The low-z imaging used in this paper and the \textit{Hubble Space Telescope} high-z data set used by F06 differ. Possible consequences are that the surface brightness cutoffs imposed by the S/N$>$1 requirement could be somewhat dissimilar, or that the imaging resolves different physical distances. While it is beyond the scope of this paper to make a thorough exploration of the surface brightness and effects of the physical resolution of images in the two samples, there are good indicators suggesting that any such difference is not very significant. 

Despite not correcting for $(1+z)^{4}$ cosmological dimming, F06 in fact found that fractional flux measurements for SN types and LGRB were not significantly different across the redshift range of their hosts from z$\sim$0.3 to z$\sim$3.  F06 also varied their surface brightness threshold by a magnitude and found this had no significant effect on their results, a step we repeated in our data set with the same outcome. Almost no SN II in this paper or ``core-collapse" SN in F06 have a zero fractional flux value despite the fact that their populations linearly follow their hosts' light distribution in both data sets, indicating that a high fraction of host light is captured in both data sets.

The sizes of galaxies in this local sample and those studied in F06 differ, which may be partly attributed to the targeting biases in local SN searches toward luminous hosts. F06 measured the absolute Johnson V-band magnitudes and absolute sizes of galaxies, $r_{80}$, the elliptical semi-major axis containing 80\% of the galactic light, using SExtractor, and found that LGRB hosts were smaller than core-collapse (CC) SN hosts: CC SN $9.62 \pm 5.58$ and LGRB $3.36 \pm 1.82$ (kpc).  We repeat these measurements for our sample: SN Ia $14.12 \pm 7.6$, SN II $13.94 \pm 7.75$, SN Ic $16.57 \pm 10.4$, and SN Ib $14.54 \pm 5.6$ (kpc).

\section{Conclusions and Summary}

SN Ic are generally found in the brightest regions of their hosts. Despite spectroscopic similarities to SN Ic,  SN Ib as well as SN II approximately track their hosts' light. A reasonable suggestion is that the brightest locations correspond to the largest star-forming regions, where the most massive stars form. These are where stars are massive enough to become SN Ic. More moderate mass stars that are produced in smaller star-forming regions may lead to most SN Ib and SN II, depending on their state of mass loss. While very massive stars can lose their outer envelope through stellar winds during a Wolf-Rayet/Luminous Blue Variable phase,  less massive stars may only be able to lose their outer H layer through stripping by a binary companion. Whether an intermediate-mass star has such a binary companion may determine if it explodes as a SN Ib or SN II.

When only light from the star-forming disk population is considered, the association of SN Ic with the brightest regions of their hosts becomes even more extreme. F06 has shown a similar result for LGRB in their irregular hosts, which lack bulges. These are presumably star-forming regions where conditions are right for forming SN Ic and, in some cases, LGRB.  The most massive stars plausibly form in these places. In the case where stellar evolution branches in one direction, perhaps because of a low abundance of heavy elements \citep{mod08}, the result may be a massive Wolf-Rayet star with no surface H or He, a compact envelope, and high angular momentum at the time of collapse that can become a LGRB (\citealt{yo05}; \citealt{wo06}). In more metal rich sites, stellar evolution takes another branch which results in a SN Ic that has similar chemistry of its atmosphere, but a less energetic core collapse. Whether it is the state of mass loss at the time of collapse or the energy release of the collapsing core that determines whether a star becomes a LGRB or a SN Ic remains to be established. That SN Ib and SN II do not have a similar preference for the brightest regions perhaps suggests that only broad-lined SN Ic share the unique explosion mechanism of LGRB. It is interesting to speculate that the distribution of broad-lined SN Ic in their hosts may have an even stronger connection to the sites of long $\gamma$-ray bursts and provide a better understanding of the violent deaths of stars. 

\acknowledgements
 
We especially thank P. Challis and M. Modjaz as well as T. Matheson, A. Fruchter, D. Mink, M. Blanton, W. Li, M. Hicken, J. Bloom, C. Blake, and S. Blondin for their help and expert advice. Research on SN at Harvard University is supported by the NSF by grant AST0606772. Additional funding for this work came through a summer grant from the Harvard College Research Program and the NSF under grant PHY05-51164 to the Kavli Institute for Theoretical Physics. Work also supported in part by the U.S. Department of Energy under contract number DE-AC02-76SF00515. 

Funding for the SDSS and SDSS-II has been provided by the Alfred P. Sloan Foundation, the Participating Institutions, the National Science Foundation, the U.S. Department of Energy, the National Aeronautics and Space Administration, the Japanese Monbukagakusho, the Max Planck Society, and the Higher Education Funding Council for England. 

The SDSS is managed by the Astrophysical Research Consortium for the Participating Institutions. The Participating Institutions are the American Museum of Natural History, Astrophysical Institute Potsdam, University of Basel, Cambridge University, Case Western Reserve University, University of Chicago, Drexel University, Fermilab, the Institute for Advanced Study, the Japan Participation Group, Johns Hopkins University, the Joint Institute for Nuclear Astrophysics, the Kavli Institute for Particle Astrophysics and Cosmology, the Korean Scientist Group, the Chinese Academy of Sciences (LAMOST), Los Alamos National Laboratory, the Max-Planck-Institute for Astronomy (MPIA), the Max-Planck-Institute for Astrophysics (MPA), New Mexico State University, Ohio State University, University of Pittsburgh, University of Portsmouth, Princeton University, the United States Naval Observatory, and the University of Washington.

\bibliography{ms}

\clearpage

\LongTables
\begin{deluxetable}{lcccccc}
\tablecaption{Measurements at SN Locations}
\tablewidth{0pt}
\tablecolumns{6}
\tablehead{\colhead{SN}&\colhead{Type}&\colhead{Morph.}&\colhead{Fractional Flux}&\colhead{\textit{g'} Surface Brightness}\\
\colhead{ }&\colhead{ }&\colhead{ }&\colhead{ }&\colhead{(mag)}}
\tablehead{\colhead{SN}&\colhead{Type}&\colhead{Morph.}&\colhead{Fractional Flux}&\colhead{Fractional Flux}&\colhead{\textit{g'} Surface Brightness}\\
\colhead{ }&\colhead{ }&\colhead{ }&\colhead{ }&\colhead{No Bulge}&\colhead{(mag)}}

\startdata

1954B&Ia&Scd&0.69&0.71&21.55&\\
1959C&Ia&SBc&0.45&...&22.05&\\
1960B&...&S0&0.16&...&23.01&\\
1960I&I*&SBc&0.33&...&22.64&\\
1960M&I&SBbc&0.42&...&21.90&\\
1960N&I&Sd&0.46&...&22.06&\\
1960R&Ia&S0/a&0.27&...&22.47&\\
1961F&IIB-L:&SBbc&0.97&0.97&20.64&\\
1961H&Ia&E&0.80&...&18.63&\\
1962A&Ia*&Sb&0.32&...&...&\\
1962B&I&Sa&0.54&...&21.12&\\
1963I&Ia&SBcd&0.79&0.79&21.75&\\
1963K&I:&S0/a&0.33&...&22.37&\\
1963M&I:&Sc&0.49&...&22.19&\\
1963P&Ia&Sc&0.18&0.18&21.77&\\
1964L&Ic&Sc&0.97&0.99&...&\\
1971G&I&SBa&0.20&...&23.11&\\
1979B&Ia&Scd&0.10&0.11&23.74&\\
1982W&Ia&S0&0.13&...&23.66&\\
1983G&Ia&S0&0.76&...&20.17&\\
1983U&Ia&SBa&0.63&0.69&21.28&\\
1984A&Ia&SBa&0.32&...&21.34&\\
1984E&IIL&Sa&0.23&0.28&22.70&\\
1984L&Ib&SBc&0.95&1.00&21.17&\\
1985B&Ia&SBa&0.36&0.40&22.53&\\
1985F&Ib/Ic&SBd&0.99&...&19.90&\\
1985G&IIP&Sbc&0.80&0.80&19.90&\\
1986A&Ia&SBc&0.72&0.76&21.10&\\
1986I&IIP&Sc&0.78&0.81&20.31&\\
1987F&IIn&Sc&0.66&0.63&21.29&\\
1987K&IIb:&Sc&0.61&0.65&20.51&\\
1987N&Ia&Sb&0.83&0.87&20.73&\\
1988Q&II&...&0.59&0.58&21.14&\\
1989A&Ia&SBbc&0.22&0.32&22.85&\\
1989E&Ib&Sc&0.41&...&22.80&\\
1989F&II&SBd&0.07&...&...&\\
1989K&II&SBab&0.03&0.06&24.10&\\
1989N&II&Sbc&0.73&0.76&21.37&\\
1990B&Ic&Sbc&0.87&0.88&20.31&\\
1990G&Ia&Sab&0.96&1.00&19.30&\\
1990H&II&Sc&0.67&0.68&20.90&\\
1990N&Ia&SBbc&0.07&0.09&23.86&\\
1991F&Ia pec&S0&0.26&...&21.57&\\
1991L&Ib/Ic&Sc&0.40&0.42&23.09&\\
1991N&Ic&SBbc&0.92&...&17.89&\\
1991S&Ia&Sab&0.20&...&23.70&\\
1991ak&Ia&SBa&0.47&0.56&22.85&\\
1991am&Ia&S&0.54&...&23.09&\\
1991bc&Ia&S0/a&0.61&...&21.24&\\
1992G&Ia&Sc&0.85&0.86&20.72&\\
1992I&II&SBbc&0.00&0.00&...&\\
1992P&Ia&Sbc&0.28&...&22.64&\\
1993G&II&IBm pec&0.28&...&22.31&\\
1993I&Ia&S0&0.00&...&25.30&\\
1993Z&Ia&Sab&0.54&...&21.17&\\
1994D&Ia&S0&0.81&...&18.95&\\
1994J&Ia&...&0.58&...&22.20&\\
1994M&Ia&E&0.26&...&23.98&\\
1994O&Ia&Sa&0.61&...&20.18&\\
1994Q&Ia&S0&0.43&...&22.00&\\
1994S&Ia&Sab&0.40&...&22.78&\\
1994W&IInP&Sbc&0.48&...&21.34&\\
1994Y&IIn&SBbc&0.64&0.69&21.63&\\
1994ae&Ia&Sc&0.18&0.18&22.69&\\
1994ak&IIn&SBa&0.23&0.28&22.81&\\
1995F&Ic&Sa&0.87&...&20.07&\\
1995H&II&Sc&0.87&0.93&21.20&\\
1995J&II&SBd&0.24&0.24&23.61&\\
1995L&Ia:&SBa&0.40&0.47&22.64&\\
1995P&Ia&...&0.66&...&21.99&\\
1995R&Ia&Sbc&0.99&0.99&20.37&\\
1995V&II&SBc&0.53&...&21.18&\\
1995al&Ia&Sbc&0.35&...&20.93&\\
1996B&II&Sbc&0.71&0.76&21.43&\\
1996V&Ia&SBa pec:&0.02&...&...&\\
1996ai&Ia&Sbc&0.80&...&19.28&\\
1996an&II&Sc&0.95&...&19.04&\\
1996aq&Ic&Scd&0.88&0.89&21.79&\\
1996bk&Ia&S0&0.55&...&19.91&\\
1996cc&II&SBc:&0.37&...&22.24&\\
1997bn&II&Scd&0.94&0.97&21.19&\\
1997co&II&Sb&0.46&0.51&21.88&\\
1997dd&IIb&Sc&0.04&...&...&\\
1997ef&Ic-bl&Sc&0.90&0.94&21.47&\\
1997ei&Ic&Sbc&0.86&0.96&21.02&\\
1998C&II&Sbc&0.26&...&23.00&\\
1998R&II&Sa&0.67&0.67&20.15&\\
1998aa&Ia&S&0.94&...&18.40&\\
1998ab&Ia pec&SBbc pec&0.41&0.45&22.08&\\
1998aq&Ia&Sb:&0.31&...&21.26&\\
1998cc&Ib&Sbc:&0.33&0.33&22.38&\\
1998cs&Ia&Sb&0.60&...&22.69&\\
1998ct&IIn&Scd:&0.93&0.93&21.21&\\
1998dk&Ia&Sc:&0.72&0.73&21.50&\\
1998dl&II&Sc&0.40&...&20.25&\\
1999Z&IIn&Sb&0.78&0.91&21.29&\\
1999aa&Ia pec&Sc&0.63&0.70&22.30&\\
1999ap&II&...&0.70&0.70&22.48&\\
1999bc&Ic&S pec&0.64&0.68&...&\\
1999bu&Ic&Sa pec&0.92&1.00&20.76&\\
1999cb&Ia&Sc&0.11&...&23.29&\\
1999cc&Ia&Sc&0.47&...&22.13&\\
1999cd&II&Sbc&0.34&0.36&22.32&\\
1999cf&Ia&SBbc&0.00&...&...&\\
1999df&II&...&0.85&...&21.77&\\
1999dg&Ia&S0&0.63&...&21.48&\\
1999eh&Ib&Sc:&0.27&0.28&22.31&\\
1999ew&II&S0/a:&0.61&...&20.93&\\
1999gj&Ia&SBbc&0.26&0.29&22.15&\\
1999gk&II&Scd&0.47&0.47&22.53&\\
1999gq&II&Sm&0.26&...&23.58&\\
2000J&II&Sbc&0.00&...&25.86&\\
2000K&Ia&S0&0.02&...&24.15&\\
2000O&Ia&S:&0.15&0.17&23.17&\\
2000bs&II&Sb&0.06&0.07&23.76&\\
2000ck&II pec&Sa&0.11&...&22.51&\\
2000cm&Ia&...&0.67&0.67&22.15&\\
2000cp&Ia&Sab&0.61&...&21.22&\\
2000cr&Ic&Sb pec&0.78&0.81&20.99&\\
2000cs&II pec&S?&0.54&0.62&22.36&\\
2000db&II&Sbc:&0.64&0.69&20.04&\\
2000df&Ia&E/S0&0.20&...&23.03&\\
2000du&II&Sb&0.43&0.52&22.06&\\
2000dv&Ib&Sb&0.55&0.58&21.16&\\
2000ez&II&SBm&0.42&0.46&21.37&\\
2000fn&Ib&Sab&0.34&0.35&21.88&\\
2001D&II&SBb&0.19&0.21&22.94&\\
2001F&Ia&Sc&0.78&...&21.91&\\
2001H&II&Scd&0.92&1.00&20.57&\\
2001J&II&SBcd:&0.68&0.70&22.28&\\
2001K&II&Sbc&0.45&0.46&21.54&\\
2001N&Ia&Sb:&0.83&...&19.98&\\
2001R&II&Sbc:&0.23&0.24&22.95&\\
2001V&Ia&Sb&0.07&...&23.47&\\
2001ab&II&SBbc:&0.23&0.24&22.52&\\
2001ad&IIb&Sc&0.23&...&23.80&\\
2001ae&II&SBb:&0.67&0.91&20.94&\\
2001ax&II&...&0.31&0.39&21.93&\\
2001ay&Ia&Sbc&0.10&0.11&...&\\
2001bk&II&...&0.07&...&24.56&\\
2001cg&Ia&SB0:&0.39&...&21.71&\\
2001cj&Ia&SBb&0.00&...&...&\\
2001ck&Ia&Sb&0.34&...&22.45&\\
2001cm&II&Sb&0.28&0.36&22.72&\\
2001co&Ib/Ic pec&SBb&0.24&0.26&22.47&\\
2001dq&Ic?&Sc&0.66&0.69&21.83&\\
2001em&Ic:&Sab&0.57&0.60&21.57&\\
2001fe&Ia&Sa&0.53&...&21.51&\\
2001gb&Ia&Sbc&0.25&0.25&22.59&\\
2001hg&II&Sbc&0.46&0.48&21.62&\\
2002G&Ia&SB?&0.24&...&22.22&\\
2002I&Ia&SBb:&0.36&0.45&21.92&\\
2002bf&Ia&SBb:&0.73&0.97&21.20&\\
2002bl&Ic-bl&SBb:&0.33&0.34&22.39&\\
2002bo&Ia&Sa pec&0.49&...&21.33&\\
2002bz&Ia&S:&0.73&...&21.44&\\
2002ca&II&SBab&0.36&0.44&21.93&\\
2002cg&Ic&Sb&0.84&0.99&20.91&\\
2002df&Ia&Sab&0.09&0.10&23.58&\\
2002dg&Ib&...&0.14&...&23.16&\\
2002ea&IIn&Sb&0.65&0.68&20.39&\\
2002ew&II&...&0.28&...&21.73&\\
2002ha&Ia&Sab&0.39&...&22.45&\\
2002hm&II&SBdm:&0.84&...&20.84&\\
2002hn&Ic&Sc&0.96&0.99&20.05&\\
2002ho&Ic&SBb&0.62&0.64&21.58&\\
2002ji&Ib/Ic&Sc:&0.25&0.25&21.51&\\
2003A&Ib/Ic&Sb&0.65&...&21.15&\\
2003I&Ib&S?&0.33&0.36&21.86&\\
2003J&II&SBb&0.58&0.62&21.38&\\
2003Y&Ia&S0&0.20&...&22.77&\\
2003ab&II&Scd:&0.48&0.49&22.59&\\
2003ag&Ia&SBbc&0.31&0.32&22.30&\\
2003aq&IIP&SBbc&0.58&0.59&22.37&\\
2003au&Ia&S0:&0.74&...&21.29&\\
2003bk&II&Scd?&0.68&...&20.98&\\
2003bm&Ic&Scd&0.19&0.22&23.81&\\
2003cn&II&Scd?&0.04&0.04&24.02&\\
2003cq&Ia&Sbc&0.21&0.22&22.79&\\
2003da&II&Scd:&0.41&0.43&21.88&\\
2003dg&Ib/Ic pec&Scd:&0.86&...&21.40&\\
2003du&Ia&SBdm&0.43&...&23.13&\\
2003ej&II&Scd:&0.67&0.68&22.52&\\
2003gm&II:&SBc:&0.71&0.72&22.61&\\
2003gs&Ia pec&SB0&0.65&...&20.65&\\
2003hi&II&S?&0.22&0.24&22.53&\\
2003ia&Ia&S?&0.50&...&22.36&\\
2003ic&Ia&SB0?&0.69&...&21.58&\\
2003jb&Ia&SB0&0.31&...&22.11&\\
2003je&II&Sab&0.35&0.39&22.69&\\
2003jz&Ia&S:&0.64&...&20.84&\\
2003ky&II&Sa&0.21&0.28&21.55&\\
2003ld&II&S?&0.78&0.78&20.52&\\
2004C&Ic&SBc&0.69&...&20.67&\\
2004G&II&Scd&0.34&0.35&22.82&\\
2004H&Ia&E&0.65&...&21.17&\\
2004I&II&SBb&0.71&0.71&20.50&\\
2004T&II&Sb&0.56&0.56&22.35&\\
2004W&Ia&E&0.25&...&22.48&\\
2004Z&II&...&0.20&0.23&23.14&\\
2004ak&II&Sbc&0.12&...&23.26&\\
2004ap&Ia&...&0.08&...&...&\\
2004aq&II&Sb&0.26&0.28&22.96&\\
2004at&Ia&Sb&0.01&...&24.54&\\
2004bj&Ia&E: pec&0.39&...&22.95&\\
2004bn&II&S?&0.46&0.51&20.89&\\
2004cm&II&...&1.00&...&20.53&\\
2004cn&Ia&...&0.94&...&22.27&\\
2004cq&Ia&Scd:&1.00&...&20.23&\\
2004ct&Ia&S?&0.57&0.61&21.57&\\
2004dg&II&Sb&0.42&0.46&21.63&\\
2004di&Ia&S0&0.10&...&23.74&\\
2004dt&Ia&SBa&0.77&...&21.73&\\
2004dv&II&SBb&0.01&0.01&24.87&\\
2004eb&II:&...&0.35&...&20.73&\\
2004el&II&Sc&0.14&...&23.17&\\
2004es&II&Sbc&0.39&0.40&22.85&\\
2004ey&Ia&SBc:&0.66&0.69&22.12&\\
2004ez&II&Sc&0.14&0.14&23.36&\\
2004gk&Ic:&Sdm:&0.75&...&21.67&\\
2004gl&Ia&...&0.65&...&22.42&\\
2004gu&Ia&...&0.54&...&22.38&\\
2004gv&Ib/Ic:&S0/a:&0.66&0.75&21.61&\\
2004hu&Ia&...&0.49&...&22.09&\\
2004ib&Ic-bl&...&0.85&0.85&...&\\
2004ie&Ia&...&0.22&...&23.54&\\
2005E&Ib/Ic&S0/a&nan&0.00&...&\\
2005G&Ia&Scd:&0.14&...&23.11&\\
2005H&II&S0: pec&0.63&...&18.80&\\
2005J&II&Sb&0.28&0.30&22.69&\\
2005O&Ib&SBbc&0.61&0.90&20.82&\\
2005S&Ia&Scd:&0.88&0.95&21.03&\\
2005T&II&Sbc&0.47&...&22.09&\\
2005U&IIb&...&1.00&...&18.73&\\
2005Y&II&Sa&0.71&...&21.17&\\
2005ab&II&Sb&0.31&0.34&22.68&\\
2005ad&II&Sc&0.04&0.04&23.62&\\
2005au&II&Scd:&0.66&0.68&21.25&\\
2005bb&II&Sb pec&0.65&0.65&21.10&\\
2005bi&II&Sbc&0.34&0.34&22.11&\\
2005bj&Ic:&...&0.20&0.22&...&\\
2005bw&II&SBbc&0.45&...&21.99&\\
2005cl&IIn&SBb&0.35&0.40&22.55&\\
2005cr&Ia&...&0.90&...&20.19&\\
2005dh&Ia&S?&0.16&...&22.21&\\
2005eo&Ic&Sbc&0.39&0.41&...&\\
2005hm&Ib&...&0.95&0.95&...&\\
2005kl&Ic&Sa&0.70&...&...&\\
2005mf&Ic&Scd&0.61&0.67&...&\\
2005mn&Ib&...&0.84&0.84&...&\\
2005nb&Ic&SBd pec&0.77&0.84&...&\\
2006cb&Ib&Sbc&0.39&0.96&...&\\
2006ck&Ic&Sd&0.89&0.96&...&\\
2006fo&Ic&...&0.58&0.60&...&\\
2006jc&Ib/Ic pec&SBbc&0.44&0.44&...&\\
2006jo&Ib&...&0.27&0.30&23.05&\\
2006lc&Ib/Ic&S0/a pec:&0.48&0.48&...&\\
2006lv&Ib/Ic&Sbc&0.53&0.55&...&\\
2006nx&Ib/Ic&...&0.01&...&...&\\
2006qk&Ic-bl&...&0.90&0.90&20.98&\\
2007I&Ic-bl&...&0.68&0.68&22.56&\\
2007ag&Ib&Scd:&0.65&...&...&\\
2007bg&Ic&...&0.00&0.00&...&\\

\enddata

\tablecomments{We list data values for SN Ib/Ic and other ambiguously typed SN as well as SN classified from light curves, even though they are not included in plots in the main text. A colon (``:") or question mark (``?") indicates some uncertainty in the classification while an asterisk (``*") denotes classification made using a light curve. Each SN type plotted in this Paper includes all subtypes and SN with peculiar (``pec") designations. These subtypes include broad-lined (``Ib-bl") for SN Ic and linear (``IIL"), plateau (``IIP"), with narrow emission lines (``IIn"), and the transitional type (``IIb") for SN II. } 
\label{tab:data}
\end{deluxetable}

\end{document}